\documentclass[nato,numreferences]{crckbked}

\begin{document}

\begin{opening}

\title{Can the Cabibbo mixing originate from \\
        noncommutative extra dimensions?}

\author{Alexandre \surname{Gavrilik}
\thanks{\email{omgavr@bitp.kiev.ua}}}

\institute{Bogolyubov Institute for Theoretical Physics, Kiev,
                                                         Ukraine}

\runningauthor{Alexandre Gavrilik}
\runningtitle{Cabibbo Mixing from Noncommutative Extra Dimensions}

\end{opening}

Abstract.
{Treating hadronic flavor symmetries with quantum algebras
$U_q(su_n)$ leads to interesting consequences such as: new
mass sum rules for hadrons $1^-$, ${\frac12}^+$, ${\frac32}^+$
of improved accuracy; possibility to label different flavors
topologically - by torus winding number;  properly fixed
deformation parameter $q$ in case of baryons is linked
in a simplest way to the Cabibbo angle
$\theta_{\rm\scriptscriptstyle{C}}$, that suggests for
$\theta_{\rm\scriptscriptstyle{C}}$ the exact value $\frac{\pi}{14}$.
In this connection, we discuss the possibility that this angle
and the Cabibbo mixing as a whole take its origin in
noncommutativity of some additional, with regard to 3+1,
space-time dimensions. }


\section{Introduction}

The problem of fermion flavors, mixings and masses (see e.g.,  \cite{gavr1})
belongs to most puzzling ones in particle physics. The Cabibbo mixing
first introduced for three lightest flavors in the context of weak
decays \cite{gavr2} involves the angle $\theta_{\rm\scriptscriptstyle{C}}$.
Importance of this concept was further confirmed after its generalization to
mixing of 3 families  \cite{gavr3}.  Due to Wolfenstein parametrization
\cite{gavr4} of CKM matrix, the Cabibbo angle now plays a prominent role: not
only CKM matrix elements $V_{ij}$, but also the quark (and even lepton)
mass ratios are often expressed as powers of small parameter
$\lambda =\sin\theta_{\rm\scriptscriptstyle{C}}\approx 0.22$.  No doubt,
it is necessary to know the value of $\lambda$ as precise as possible.
In this respect, the main bonus of our approach to flavor symmetries, based
on quantum algebras, is that it suggests {\it theoretically motivated} exact
value for $\theta_{\rm\scriptscriptstyle{C}}$, namely,
$\theta_{\rm\scriptscriptstyle{C}}\!=\!\frac{\pi}{14}$. As further
implication, it leads us to a conjecture of possible
noncommutative-geometric origin of the Cabibbo mixing, and our aim here is
to argue this may indeed be the case.
Below, when treating baryon masses, we restrict ourselves with 4 flavors
including u-, d-, s-, and c- quarks.  Basic tool of the aproach used is
the representation theory of quantum algebras  \cite{gavr5} $U_q(su_n)$
adopted, instead of conventional $SU(n)$, to describe flavor symmetries
classifying hadrons into multiplets.

\section{Vector meson masses: $q$-deformation replaces (singlet) mixing}
We use\footnote{
For more details concerning this approach see refs.
\cite{gavr6,gavr7,gavr13}.}
Gelfand-Tsetlin basis vectors for meson states from $(n^2\!-\!1)$-plet of
'$n$-flavor' $U_q(u_n)$ embedded into $\{(n\!+\!1)^2\!-\!1\}$-plet of
'dynamical' $U_q(u_{n+1});\ $ construct mass operator $\hat {M_n}$
invariant under the 'isospin+hypercharge' $q$-algebra $U_q(u_2)$ from
generators of dynamical algebra $U_q(u_{n+1})\ $
(e.g., $\hat {M_3} = M_0{\bf 1}+{\gamma}_3 A_{34}A_{43}
+{\delta}_3 A_{43}A_{34}$);  calculate the expressions for masses
$m_{V_i}\equiv\langle V_i|\hat{M_3}|V_i\rangle $ -
these involve $M_0$, symmetry breaking parameters $\gamma_3, {\delta}_3$,
and the $q$-parameter. In particular, for $n=3$ we obtain
\vspace{-1.0mm}
\begin{equation}
m_{\rho}=M_0,  \hspace{9mm}  m_{K^*}=M_0 -{\gamma}_3,
            \hspace{10mm}
m_{{\omega}_8}= M_0 - 2\ \frac{[2]_q}{[3]_q}\ {\gamma}_3,   \label{(1)}
\vspace{-0.6mm}
\end{equation}
where $[x]_q \equiv \frac{q^x - q^{-x}}{q - q^{-1}}$ is the $q$-number
that 'deforms' a number $x$ and, to have equal masses for particles and
their anti's, $\delta_3\!=\!\gamma_3$ was set. $q$-Dependence appears only
in the mass of ${\omega }_8$ (isosinglet in $U_q(su_3)$-octet). Excluding
$M_0, \gamma_3$, the $q$-analog of Gell--Mann - Okubo (GMO)
relation is \cite{gavr8} :
\vspace{-1.0mm}
\begin{equation}
m_{{\omega }_8} + \biggl (2 {\frac{[2]_q}{[3]_q}}
- 1\biggr ) m_{\rho } = 2 {\frac{[2]_q}{[3]_q}} m_{K^*}\ .  \label{(2)}
\end{equation}
In the limit $q\!=\!1$ (then, $\frac{[2]_q}{[3]_q}=\frac23$), 
this reduces to
usual GMO formula $3m_{{\omega}_8}\!+\! m_{\rho} = 4 m_{K^*}$
which needs singlet mixing      \cite{gavr9}.  However, it also yields
\begin{equation} m_{{\omega}_8} + m_{\rho} = 2 m_{K^*}  \hspace{12mm}
    \ {\rm if}\ \ \ \ \ q=e^{i\pi /5} \ \
\left( {\rm then,} \hspace{2mm} [2]_q=[3]_q \right) .         \label{(3)}
\end{equation}
With $m_{{\omega}_8}\equiv m_{\phi}$, and no mixing, eq.(3) coincides
with nonet mass formula of Okubo  \cite{gavr10}
agreeing {\it ideally} with data \cite{gavr11}.

For $3 < n \le 6$ mass operator is constructed analogously. Again,
calculations show: only singlets
${\omega}_{15}$, ${\omega}_{24}$, ${\omega}_{35}$ of $(n^2-1)$-plets
of $U_q(u_n)$ contain $q$-dependence. As result, we get
the $q$-deformed mass relations \cite{gavr8,gavr6,gavr7}:
\vspace{-1.4mm}
\begin{equation}
[n]_{(q)}\  m_{\omega_{n^2-1}} + (b_{n;q}+2n-4) \ m_{\rho} =
2\ m_{D_n^*} + (c_{n;q} + 2) \sum_{r=3}^{n-1}m_{D_r^*} ,   \label{(4)}
\end{equation}
\vspace{-1.8mm}
\[
b_{n;q}\equiv n \ c_{n;q} -6\ [n]^2_{(q)} +
\biggl (\frac{24}{[2]_q}-1\biggr )[n]_{(q)} ,        \hspace{8mm}
 c_{n;q}\equiv 2 \ [n]^2_{(q)}-\frac{8}{[2]_q}[n]_{(q)} ,
\]
\vspace{0.2mm}
where $[n]_{(q)} \equiv [n]_q/[n\!-\!1]_q$. Then, natural fixation by
setting $[n]_q=[n\!-\!1]_q$, $n=4,5,6,$ leads to the higher analogs of
Okubo's nonet sum rule:
\begin{equation}
m_{{\omega}_{15}} + (5 - 8/[2]_{q_4}) m_{\rho} =
 2\ m_{D^*} + (4 - 8/[2]_{q_4}) m_{K^*}  \hspace{36mm}   \label{(5)}
\end{equation}
\vspace{-2.2mm}
\begin{equation}
m_{{\omega}_{24}} + (9 - 16/[2]_{q_5}) m_{\rho} =
2\ m_{{D_b}^*} + (4 - 8/[2]_{q_5})(m_{D^*}
                             + m_{K^*})  \hspace{18mm}   \label{(6)}
\end{equation}
\vspace{-1.8mm}
\begin{equation}
 m_{{\omega}_{35}} + (13 - 24/[2]_{q_6}) m_{\rho} =
2\ m_{{D_t}^*}
+ (4 - 8/[2]_{q_6})(m_{{D_b}^*} + m_{D^*} + m_{K^*}).    \label{(7)}
\vspace{1.4mm}
\end{equation}
Here $q_n=e^{i\pi/(2n-1)}$ are
the values that solve eqns. $\ [n]_q-[n\!-\!1]_q=0$.
Like in the case with $m_{{\omega}_8}\equiv m_{\phi},$ it is meant
in (5)-(7) that $J/\psi $ is put in place of ${\omega }_{15},$
$\Upsilon$ in place of ${\omega }_{24},$ toponium in place of
${\omega }_{35}$ (i.e., \underline{no mixing}!).

The $q$-polynomials $[n]_q-[n\!-\!1]_q$ have a topological meaning.

\section{Torus knots and topological labelling of flavors }
Polynomials $[n]_q\!-\![n\!-\!1]_q\equiv P_n(q),$ by their roots,
reduce $q$-analogs (2), (4) to realistic mass sum rules (MSR)
(3), (5)-(7). And, due to property
$(i)\ P_n(q)=P_n(q^{-1}),\ (ii)\ P_n(1)=1,$ they coincide \cite{gavr8,gavr7}
with such knot invariants as Alexander polynomials
$\Delta (q)\{(2n\!-\!1)_1\}$ of $(2n-1)_1$-torus knots. E.g.,
\vspace{-2.4mm}
\[  \ \ \ \
[3]_q-[2]_q=q^2+q^{-2}-q-q^{-1}+1\equiv \Delta (q)\{ 5_1\} ,
                                       \hskip 20mm
\]
\[  \ \
[4]_q-[3]_q=
q^3+q^{-3}-q^2-q^{-2}+q+q^{-1}-1\equiv \Delta (q)\{ 7_1\}
\vspace{0.4mm}
\]
correspond to the $5_1$- and $7_1$-knots. Since the $q$-deuce in (4)
can be linked to the trefoil (or $3_1$-) knot:
$[2]_q -1=q+q^{-1}-1\equiv\Delta (q)\{3_1\}$,
{\it all the} $q$-{\it dependence} in masses of $\omega_{n^2-1}$
and in coefficients in (2),(4) is expressible through
Alexander polynomials. Namely,
$
\frac{[3]_q}{[2]_q}=1+\frac{\Delta \{5_1\}}{[2]_q}=
1+\frac{\Delta \{5_1\}}{\Delta \{3_1\} +1},  
$
\begin{equation}
\vspace{0.2mm}
\frac{[n]_q}{[n-1]_q }=
1+\frac{\Delta \{ (2n-1)_1\}}{[n-1]_q}=
1+\frac{\Delta \{ (2n-1)_1\} }{1+ \sum_{r=2}^{n-1} \Delta \{(2r-1)_1\} },  
            \ \ \   n=4,5,6.                          \label{(8)}
\end{equation}
The values $q_n$ are thus roots of respective Alexander
polynomials. For each $n$, the \underline{'senior'} (numerator)
polynomial in $\frac{[3]_q}{[2]_q}$ and (8) is specified:
by its root, it 'singles out' the corresponding MSR from
$q$-deformed analog.

Thus, the $q$-parameter for each $n$ is fixed in a rigid way
as a root $q_n$ of $\Delta \{ (2n-1)_1\},$ contrary to the choice of $q$
by fitting in other phenomenological applications      \cite{gavr12}.
Moreover, using flavor $q$-algebras along with 'dynamical' $q$-algebras
according to $U_q(u_{n})\subset U_q(u_{n+1})$, we gain: the torus knots
$5_1,\ 7_1,\ 9_1,\ 11_1$ \ are put into correspondence  \cite{gavr6,gavr7}
with vector quarkonia $s\bar s$, $c\bar c$, $b\bar b$, and $t\bar t$
respectively. In a sense, the polynomial $P_{n}(q)\equiv [n]_q-[n\!-\!1]_q$
by its root $q(n)$ \underline{determines} the value of $q$
(deformation strength) for each $n$ and thus serves as
  {\it defining polynomial} for the MSR/quarkonium/flavor
corresponding to $n$.  Hence, the applying of $q$-algebras suggests
a possibility of {\it topological labeling of flavors}:\ fixed number $n$
corresponds to $2n\!-\!1$ overcrossings of 2-strand braids whose
closure gives these $(2n-1)_1$-torus knots. With the form $(2n\!-\!1,2)$
of same torus knots this means the correspondence
$ n \leftrightarrow w\equiv 2n-1,$ $\ w$ being the winding number
around tube of torus (winding number around hole is 2).

\section{Defining $q$-polynomials for octet baryon mass sum rules }
Analogous scheme was applied to baryons ${\frac12}^+$ too.  Excluding
undetermined constants $M_0,\alpha$, $\beta$ from final obtained expressions
for $M_N$, $M_\Xi$, $M_\Lambda$, $M_\Sigma$ leads to the $q$-deformed
mass relations (MRs) of the form       \cite{gavr6,gavr7,gavr13}
\[
[2]M_N+\frac{[2]}{[2]-1}M_{\Xi }=[3]M_{\Lambda }
+ \biggl ( \frac{[2]^2}{[2]-1}-[3]  \biggr )
          M_{\Sigma }     \hspace{14mm}
\]
\vspace{-3mm}
\begin{equation}   \hspace{48mm}
  +\frac{A_q}{B_q}\Bigl( M_{\Xi } + [2] M_N -
   [2]M_{\Sigma } - M_{\Lambda } \Bigr)                   \label{(9)}
 \end{equation}
where $A_q$ and $B_q$ are certain polynomials of $[2]_q$ with
non-overlapping sets of zeros. It is important that different dynamical
representations produce differing pairs $A_q$, $B_q$.  Any $A_q$
possesses the factor $([2]_q-2)$ and thus the 'classical' zero $q=1$.
In the limit $q=1$ each $q$-deformed mass relation reduces to
the standard GMO sum rule $M_N+M_\Xi=\frac12 M_\Sigma+\frac32 M_\Lambda$
for octet baryons (its accuracy is $0.58\%$).
At some values of $q$ which are zeros of particular $A_q$ other than
$q=1$, we obtain MSRs which hold with better accuracy than the GMO one.
The two new MSRs
\begin{equation}
{q=e^{{\rm i}\pi/6}} \ \ \   {\Rightarrow}  \ \ \ \ \
M_N+\frac{1\!+\!\sqrt{3}}{2}M_\Xi=\frac{2}{\sqrt{3}}M_\Lambda+
\frac{9\!-\!\sqrt{3}}{6}M_\Sigma\   \hspace{5mm} (0.22\%)
                                                        \label{(10)}
\end{equation}
\vspace{-3mm}
\begin{equation}
{q=e^{{\rm i}\pi/7}} \ \ \   {\Rightarrow}  \ \ \ \ \
M_N+\frac{1}{[2]_{q_7}\!-\!1}M_{\Xi}=
\frac{1}{[2]_{q_7}\!-\!1}M_{\Lambda}+M_{\Sigma}
\hspace{5.2mm}  (0.07\%)
                                                        \label{(11)}
\end{equation}
result [6,7,13] from two different dynamical representations
$D^{(1)}$ and $D^{(2)}$ whose respective polynomials
$A_q^{(1)}$ and $A_q^{(2)}$ possess zeros $q=e^{{\rm i}\pi/6}$
and $q=e^{{\rm i}\pi/7}$.  The choise with $q=e^{{\rm i}\pi/7}$ turns out
to be the best possible one.\footnote{
In sec. 8 we argue that this value of $q$ is linked to the Cabibbo angle:
$\theta_{\bf 8}\!\equiv\!\frac{\pi}{7}\!=\!2\theta_{\rm\scriptscriptstyle{C}}$.}

The sum rule (10) was first derived \cite{gavr6} from a specific
dynamical representation (irrep) $D^{(1)}$ of $U_q(u_{4,1})$.
However, the 'compact'
dynamical $U_q(u_5)$ is equally well suited. Among the admissible
dynamical irreps there exist an entire series  of irreps
(numbered by integer $m$, $6\le m<\infty $) which produce
the corresponding infinite set of  MSRs:
\begin{equation}
M_N+\frac{1}{[2]_{q_m}-1}M_{\Xi }=
\frac{[3]_{q_m}}{[2]_{q_m}}M_{\Lambda }
+\Bigl (\frac{[2]_{q_m}}{[2]_{q_m}-1}-
\frac{[3]_{q_m}}{[2]_{q_m}}\Bigr )M_{\Sigma }          \label{(12)}
\end{equation}
with $q_m=e^{{\rm i}\pi/m}$.
{\em Each of these shows better agreement with data} than
the classical GMO one. Few of them, including
the MSRs (10), (11) and the 'classical' GMO which corresponds
to $q_{\infty}=1$, are shown in the table.
\begin{center}
\begin{tabular}{|c|c|c|}
\hline
$\theta=\frac{\pi}{m}$  &  (RHS$-$LHS),\  $MeV$  &
       $\frac{\left| {\rm RHS}-{\rm LHS}\right|}{{\rm RHS} }, \%$ \\
\hline
$\pi /\infty$ &  26.2    &   0.58 \\
$\pi /30$    &   25.42   &   0.56 \\
$\pi /12$    &   20.2    &   0.44 \\
$\pi /8$     &   10.39   &   0.23 \\
$\pi /7$     &    3.26   &   0.07  \\
$\pi /6$     &  -10.47   &   0.22  \\
\hline
\end{tabular}
\end{center}
Comparing (12) with (9) shows that the vanishing of $\frac{A_q}{B_q}$
is crucial for obtaining this discrete set of MSRs and for
providing a kind of 'discrete fitting'. Thus, $A_q$ serves as
{\it defining} polynomial for the corresponding MSR.

Since $[2]_{q_7}=q_7+\frac{1}{q_7}=2\cos\frac\pi7$, the MSR (11)
takes the equivalent form
\begin{equation}
 M_{\Xi}-M_{N}+M_{\Sigma}-M_{\Lambda} =
(2\cos \frac\pi 7)(M_{\Sigma}-M_{N})                      \label{(13)}
\end{equation}
which exhibits some similarity with decuplet mass formula given below.

\section{Decuplet baryons: universal $q$-deformed mass relation}
In the case of $SU(3)$-decuplet baryons ${\frac32}^+$, the conventional
1st order symmetry breaking yields   \cite{gavr9} equal spacing rule (ESR)
for isoplet members in ${\bf 10}$-plet. Empirical data show for
$M_{\Sigma^*}-M_{\Delta},$ $ M_{\Xi^*}-M_{\Sigma^*}$ and
$M_{\Omega}-M_{\Xi^*}$ noticeable deviation from ESR:
$152.6~MeV\leftrightarrow 148.8~MeV\leftrightarrow 139.0~MeV$.
Use of the $q$-algebras $U_q(su_n)$ instead of $SU(n)$ provides 
natural improvement.  From evaluations of decuplet masses in two 
distinct particular irreps of the dynamical algebra $U_q(u_{4,1})$,
the $q$-deformed mass relation
\begin{equation}
(1/[2]_q)(M_{\Sigma^*}-M_{\Delta}+M_{\Omega}-M_{\Xi^*})
= M_{\Xi^*}-M_{\Sigma^*},
          \hspace{8mm} [2]_q\equiv q+q^{-1},                 \label{(14)}
\end{equation}
was derived   \cite{gavr14}. As proven there, this mass relation is
\underline{universal} - it results from each admissible irrep
(which contains
$U_q(su_3)$-decuplet embedded in {\em 20}-plet of $U_q(su_4)$) of
the dynamical $U_q(u_{4,1})$. With empirical masses   \cite{gavr11},
the formula (14) is successfull if $[2]_q\simeq 1.96$. {\it Pure phase}
$q=e^{i\theta}$ (or $[2]_q=2\cos\theta$) with
$\theta=\theta_{\bf 10}\simeq\frac{\pi }{14}$ provides excellent
agreement with data (below, we argue that
$\theta_{\bf 10}=\theta_{\rm\scriptscriptstyle{C}}$).
Notice a similarity of eq.(14) with the MR
\begin{equation}
(1/2)(M_{\Sigma^{*}}-M_{\Delta}+M_{\Omega}-M_{\Xi^{*}})
= M_{\Xi^{*}}-M_{\Sigma^{*}}                             \label{(15)}
\end{equation}
obtained earlier in diverse contexts   \cite{gavr15}:
by tensor method, in additive quark model with general pair interaction,
in a diquark--quark model, in modern chiral perturbation theory.
Such model-independence of (15) stems because each of these approaches
\underline{accounts 1st and 2nd order} of $SU(3)$-breaking.

The $q$-deformed MSR (14) is universal even in a wider sense: it results
from admissible irreps (containing $U_q(su_4)\ $  {\em 20}-plet)
of both $U_q(su_{4,1})$ and the 'compact' dynamical $U_q(su_5)$.
Say, within a dynamical irrep $\{ 4 0 0 0 \}$ of $U_q(su_5)$
calculation yields:
$M_{\Delta}=M_{\bf 10} +  \beta ,\ $
$M_{\Sigma^*}=M_{\bf 10} + [2] \beta +  \alpha ,\ $
$M_{\Xi^*}=M_{\bf 10} + [3] \beta + [2] \alpha ,\ $
$M_{\Omega}=M_{\bf 10} + [4] \beta + [3] \alpha ,\ $
from which (14) stems. On the other hand, these four masses
can be comprised by single formula
\begin{equation}
M_{D_i} = M\bigl( Y(D_i) \bigr ) =
M_{\bf 10} + {\alpha} [1\!-\!Y(D_i)]+{\beta} [2\!-\!Y(D_i)]   \label{(16)}
\end{equation}
with explicit dependence on $Y$ (hypercharge).
If $q=1$, this reduces to $M_{D_i} = \tilde{M}_{\bf 10} + a~Y\!(D_i)$,
i.e., {\it linear dependence on hypercharge} $Y$
(or strang\-eness) where $ a = - \alpha - \beta ,\ $
$\tilde{M}_{\bf 10} = M_{\bf 10} + \alpha + 2 \beta.$

\section{Nonpolynomial $SU(3)$-breaking effects in baryon masses}
Formula (16) {\it involves highly nonlinear dependence} of mass on
hypercharge (it is $Y$ that causes $SU(3)$-breaking for decuplet).
Since for $q$-number $[N]$ we have
$[N]=q^{N-1}+q^{N-3}+\ldots +q^{-N+3}+q^{-N+1}$ ($N$ terms)
this shows exponential $Y$-dependence of masses.
Such high nonlinearity makes (14) and (16) radically different from
the abovementioned result (15) of traditional treatment that accounts
for effects linear and quadratic in $Y$.

For octet baryon masses, high nonlinearity ({\em nonpolynomiality})
in $SU(3)$-breaking effectively accounted by the model was demonstrated
in   \cite{gavr13}.
For this, the expressions for (isoplet members of) octet masses with
explicit dependence on hypercharge $Y$ and isospin $I$, through $I(I+1)$,
are used. The typical matrix element ($\mu_1,\mu_2$ are functions of
irrep labels $m_{\scriptscriptstyle{15}},m_{\scriptscriptstyle{55}}$):
\[          \hspace{-1.0mm}
\langle B_i | A_{34} A_{45} A_{54} A_{43} | B_i \rangle
= [2]^{-1}[3]^{-1}
\Bigl({[Y/2][Y/2\!+\!1]-[I][I\!+\!1]}\Bigr)
\ \mu_1(m_{15},m_{55})
\]
\[
-[2]^{-1}[5]^{-1}
\Bigl({[Y/2-1][Y/2-2]-[I][I+1]}\Bigr)
\ \mu_2(m_{15},m_{55}),
\]
contributing to octet baryon masses, illustrates the dependence.
From definition of $q$-bracket $[n]=\frac{\sin(nh)}{\sin(h)}$,
$ q\!=\!\exp({\rm i}h)$, it is clearly seen that baryon masses depend on
hypercharge $Y$ and isospin $I$ (hence, on $SU(3)$-breaking effects) in
highly nonlinear - {\it nonpolynomial} - fashion.

The ability to take into account highly nontrivial
symmetry breaking effects by applying $q$-analogs $U_q(su_n)$ of 
usual flavor symmetries is much alike the fact 
demonstrated in \cite{gavr16} that,
by exploiting appropriate {\em free} $q$-deformed
structure one is able to efficiently study the properties of
(undeformed) quantum-mechanical systems with complicated interactions.

\section{To use or not to use the Hopf-algebra structure}
An alternative, as regards (9),  version of $q$-deformed analog
can be derived   \cite{gavr13}  using for the symmetry breaking part
of mass operator a component of $q$-{\em tensor operator} -
this clearly implies   \cite{gavr17}  the Hopf algebra structure
(comultiplication, antipode)  of the $U_q(su_n)$ quantum algebras.
Let us briefly discuss such version.
 We use $q$-tensor operators $(V_1,V_2,V_3)$ resp.
$(V_{\bar{1}},V_{\bar{2}},V_{\bar{3}})$ formed from elements of
$U_q(su_4)$ and transforming as ${{\bf 3}}$ resp. ${{\bf 3}^*}$
under the adjoint action of $U_q(su_3)$. With $H_1, H_2$ as Cartan
elements and with notation $[X,Y]_q\equiv XY-qYX$,
the components $(V_1,V_2,V_3)$ read
\[
 V_1=[E_1^+,[E_2^+,E_3^+]_q]_q q^{-H_1/3-H_2/6} ,\ \
V_2=[E_2^+,E_3^+]_q q^{H_1/6-H_2/6} ,\ \
\]
\begin{equation}
V_3=E_3^+ q^{H_1/6+H_2/3} ,                             \label{17}
\end{equation}
and similarly for $(V_{\bar{1}},V_{\bar{2}},V_{\bar{3}})$
(see \cite{gavr13}), of which we here only give
\begin{equation}
 V_{\bar{3}}=q^{H_1/6+H_2/3} E_3^- .                    \label{18}
\end{equation}
Clearly, $U_q(su_3)$ is broken to $U_q(su_2)$.
Like in the nondeformed case of $su(3)$ broken to its isospin
subalgebra $su(2)$, the form of mass operator is
\begin{equation}
\hat M = \hat M_0+ \hat M_8                              \label{19}
\end{equation}
where $\hat{M}_0$ is $U_q(su_3)$-invariant and $\hat{M}_8$ transforms
as $I\!=\!0,Y\!=\!0$ component of tensor operator of ${\bf 8}$-irrep
of $U_q(su_3).$   If $|B_i\rangle$ is a basis vector of carrier space of
${\bf 8}$ which corresponds to some baryon $B_i$,
the mass of $B_i$ is given by $M_{B_i}=\langle B_i|\hat{M}|B_i \rangle$.
The irrep ${\bf 8}$ occurs twice in the decomposition
of ${{\bf 8}}\otimes{{\bf 8}}$.
This, and the Wigner-Eckart theorem for $U_q(su_n)$  \cite{gavr18}
applied to $q$-tensor operators under irrep ${\bf 8}$ of $U_q(su_3)$,
lead to the mass operator of the form
$ \hat M = M_0 {\bf 1} + \alpha V_8^{(1)} + \beta V_8^{(2)} $
and thus to
\begin{equation}
M_{B_i} =
\langle B_i |(M_0 {\bf 1} + \alpha V_8^{(1)} + \beta V_8^{(2)})
| B_i \rangle                                               \label{20}
\end{equation}
where $V_8^{(1)}$ and $V_8^{(2)}$ are two dictinct tensor
operators which both transform as $I\!=\!0, Y\!=\!0$ component of
irrep ${\bf 8}$ of $U_q(su_3)$; $M_0,$ $\alpha$, $\beta$ - undetermined
constants depending on details of dynamics.  From
${{\bf 3}}\otimes {{\bf 3}}^*\!=\!{{\bf 1}}\oplus {{\bf 8}},$
${{\bf 3}}^*\otimes {{\bf 3}}\!=\!{{\bf 1}}\oplus {{\bf 8}}$
it is seen that the operators $V_{3}V_{\bar{3}}$ and $V_{\bar{3}}V_{3}$
from (17),(18) are just the isosinglets needed in eq.(20). As result,
mass operator in (20) with redefined $M_0,\alpha,\beta$ is
$ \hat M = M_0 {\bf 1} +\alpha V_3 V_{\bar{3}} + \beta V_{\bar{3}} V_3$,
or
\vspace{-1mm}
\begin{equation}
\hat M = M_0 {\bf 1}
 +\alpha E_3^+ E_3^- q^Y + \beta E_3^- E_3^+ q^Y               \label{21}
\end{equation}
where $Y=(H_1+2H_2)/3$ is hypercharge.
Matrix elements (20) with $\hat M$ from (21) are evaluated by
embedding ${\bf 8}$ in a particular representation of $U_q(su_4).$
Say, if one takes the adjoint ${\bf 15}$ of $U_q(su_4)$,
the evaluation of baryon masses yields:
$ M_N=M_0+\beta q ,\ M_\Sigma= M_0 ,\  M_\Lambda =M_0 +
  \frac{[2]}{[3]} (\alpha+\beta) ,\ M_\Xi=M_0+\alpha q^{-1}$.
Excluding $M_0,\alpha$ and $\beta$, we finally obtain
\begin{equation}
[3] M_\Lambda + M_\Sigma=[2] (q^{-1} M_N+q M_\Xi) .            \label{22}
\end{equation}
This alternative $q$-analog of octet mass relation looks much
simpler than the former $q$-analog (9).
This same $q$-relation (22) results from embedding
${\bf 8}$ in any other admissible dynamical representation.
What concerns empirical validity \cite{gavr11} of (22),
there is no other way to fix the $q$-parameter as by usual fitting
(for each of the values $q_{1,2}=\pm 1.035$,
$q_{3,4}=\pm 0.903 \sqrt{-1}$, the $q$-MR (22) indeed holds
within experimental uncertainty).
This is in sharp contrast with the $q$-analogs (9) for which there
exists an appealing possibility to fix $q$ in a rigid way by zeros of
relevant polynomial $A_q$.

Summarizing we should stress that, although the use of Hopf-algebra
structure leads to simple and mathematically appealing result eq.(22),
from the physical (phenomenological) viewpoint the version (9) of
$q$-analog obtained by applying only the tools of representation theory
of quantum algebras and not strictly $q$-covariant symmetry breaking part
in mass operator, provides much more interesting results. Among these is
the degeneracy lifting and the possibility to choose among a variety
of dynamical representations, defining polynomials and, thus, within
discrete set of viable mass sum rules. That led us to the best MSR (11)
(or (13)) for octet baryons.

\section{On the connection: deformation parameter $\leftrightarrow$
 Cabibbo angle}
In 3-flavor case of vector mesons, the deformation angle $\frac{\pi}{5}$
that determines $\phi$-meson in (3) coincides remarkably with
$\omega$-$\phi$ mixing angle (known    \cite{gavr11}
to be $\theta_{\omega\phi}=36^{\circ}$) of traditional $SU(3)$-based
scheme. In other words, the concept of $q$-deformed flavor symmetries
{\em is closely related} with the issue of singlet mixing.

For pseudoscalar (PS) mesons, the generalization \cite{gavr19}
of GMO-formula
\begin{equation}
f_\pi^2 m_\pi^2 + 3 f_\eta^2 m_\eta^2 = 4 f_K^2 m_K^2
 \ \ \ \ \ \ \ \  \hbox{with} \ \ \ \ \ \ \
1/{f^{2}_\pi} + {3}/{f^{2}_\eta} = {4}/{f^{2}_K},
\end{equation}                                               \label{23}
involves decay constants as coefficients.
Presented in the equivalent form\footnote{
Note that having used the additional constraint in (23) we are led
to the single dimensionless quantity $\frac{f_K}{f_\pi}$ involved
in the multipliers of masses.}
\begin{equation}
m_\pi^2 + \frac{ 9 {f_K^2}/{f_\pi^2} }
   {4-{f_K^2}/{f_\pi^2} } m_\eta^2
= 4 \frac{f_K^2}{f_\pi^2} m_K^2 ,                             \label{24}
\end{equation}
it is to be compared with our $q$-{\it analog} (2) of GMO
rewritten for PS mesons (with masses squared), namely
\begin{equation}
m_\pi^2 + \frac{[3]}{2 [2] - [3]} m_{\eta_8}^2
= \frac{2 [2]}{2 [2] - [3]} m_K^2 .                       \label{25}
\end{equation}
Without singlet mixing, it is satisfied for (the mass of)
{\it physical} $\eta$-meson put instead of $\eta_8$
{\em at properly fixed} $q=q_{\scriptscriptstyle\rm{PS}}$,
 and just this is meant below.

The two generalizations (24) resp. (25) yield the standard GMO
mass formula in the corresponding limit of single parameter,
$\frac{f_K}{f_\pi}\to 1$ resp. $q\to 1$. Moreover, the following
identification is valid:
\begin{equation}
\frac{f_K^2}{f_\pi^2} \leftarrow\!\rightarrow
\frac{\frac12 [2]}{2 [2] - [3]},
\hspace{18mm}
 \frac{ 3 {f_K^2}/{f_\pi^2} }{4-{f_K^2}/{f_\pi^2} }
\leftarrow\!\rightarrow \frac{\frac13 [3]}{2 [2] - [3]},      \label{(26)}
\end{equation}
from which, using $[3]_q=[2]_q^2-1$, we get
\begin{equation}
[2]_{\pm}=1-{\xi}_{\pi,K} \pm\sqrt{\bigl(1-{\xi}_{\pi,K}\bigr)^2 + 1 }\ ,
\ \ \ \ \ \ {\xi}_{\pi,K}\equiv({4 f^2_K/f^2_\pi})^{-1} .     \label{(27)}
\end{equation}
The ratio $f_K/f_\pi$ is related to the Cabibbo angle.
This is evident either from the formula:
   $\tan^2\theta_{\rm\scriptscriptstyle{C}} = \frac{m_\pi^2}{m_K^2}
\Bigl[\frac{f_K}{f_\pi}-\frac{m_\pi^2}{m_K^2}\Bigr]^{-1}$
(see \cite{gavr20}), or from the formula
\[
\frac{\Gamma_{K\to\mu\nu}}{\Gamma_{\pi\to\mu\nu}}
= (\tan \theta_{\rm\scriptscriptstyle{C}})^2\
\frac{f_K^2}{f_\pi^2}\ \frac{M_K}{M_\pi}
\left(
\frac{1-(M_\mu /M_K )^2}{1-(M_\mu /M_\pi )^2} \right)^2
\]
for the ratio of weak decay rates usually
applied to determine        \cite{gavr21,gavr11}
$\ f_K /f_\pi $ in terms of the Cabibbo angle,
with known empirical data on decay rates and masses.
  Thus, the value of $f_K / f_\pi$ is expressible through
$\theta_{\rm\scriptscriptstyle{C}}$.
Together with (26), (27) this implies:  within our scheme,
the (realistic value $q_{\scriptscriptstyle\rm{PS}}$ of)
{\it deformation parameter is directly connected
with the Cabibbo angle}.

Similar conclusion can be arrived at in another, more general context.
In   \cite{gavr22}, the $q$-deformed lagrangian for gauge fields of
the Weinberg - Salam (WS) model invariant under the quantum-group
valued gauge transformations was constructed. The obtained formula
\cite{gavr22}
\begin{equation}
F^0_{\mu\nu}\ =\ {\rm Tr}_q(F_{\mu\nu})\ [2(q^2 + q^{-2})]^{-1/2}
  = \ B_{\mu\nu} \cos\theta + F^3_{\mu\nu}\sin\theta ,        \label{28}
\end{equation}
\[   F^3_{\mu\nu}= \partial_\mu A^3_\nu - \partial_\nu A^3_\mu
           + {\rm i}e^{ab3} ( A^a_\mu A^b_\nu - A^a_\nu A^b_\mu )
       + [A^3_\mu, B_\nu] - [A^3_\nu , B_\mu ] ,
\]
\[ B_{\mu\nu}=\partial_{\mu} B_{\nu}-\partial_{\nu} B_{\mu} +
[B_{\mu}, B_{\nu}] + [A^a_{\mu} , A^a_{\nu} ]
\]
where
\begin{equation}
    \tan\theta = (1-q^2)/(1+q^2) ,                             \label{29}
\end{equation}
exhibits a mixing of the $U(1)$-component $B_\mu$ with nonabelian
components $A^a_\mu$ (the third one). Introducing the new potentials
$\tilde{A}_\mu = B_\mu\cos\theta + A^3_\mu \sin\theta,$
$\ Z_\mu=-B_\mu \sin\theta + A^3_\mu \cos\theta$  yields nothing but
definition of physical photon $\tilde{A}_\mu$ and $Z$-boson of WS model,
where $\theta$ coincides with the Weinberg angle,
$\theta=\theta_{\rm\scriptscriptstyle{W}}$.
Since at $\theta = 0$ the potentials $B_\mu$ and $A^3_\mu$ get
completely unmixed whereas nonzero $\theta$ (i.e., nontrivial $q$-deformation)
provides proper mixing as a characteristic feature of the WS model,
it is thus seen that the {\em weak mixing is adequately
modelled by the $q$-deformation}.
Moreover, formula analogous to (29), i.e.,
$  \tan\theta_{\rm\scriptscriptstyle{W}} = q\sqrt{ {[4]}/({[2][3]})}
            \left[1/2\right] \left[ 3/2\right],$
was obtained \cite{gavr23} within somewhat different approach
to $q$-deforming the standard model.

Hence, the $q$-deformation realizes proper mixing in the sector of
gauge fields, thus providing explicit connection between
the weak angle and the deformation parameter $q$.

On the other hand, the relation found in   \cite{gavr24}, namely
\begin{equation}
\theta_{\rm\scriptscriptstyle{W}}
         = 2 (\theta_{12} + \theta_{23} + \theta_{13}) ,         \label{30}
\end{equation}
connects $\theta_{\rm\scriptscriptstyle{W}}$ with the Cabibbo angle
$\theta_{12}\equiv \theta_{\rm\scriptscriptstyle{C}}$ (and two other
Kobayashi-Maskawa angles $\theta_{13}, \theta_{23}$;  as we deal with two
lightest families, we have to discard $\theta_{13}, \theta_{23}$).
The importance of (30) consists in that it links two apparently different
mixings: one
involved in {\it bosonic} (interaction) sector, the other in
{\it fermionic} (matter) sector of the electroweak standard model.

Combining (29) and (30) ($\theta_{23}, \theta_{13}$ omitted)
we conclude: the Cabibbo angle {\it should be connected} with
the $q$-parameter of a quantum-group (or quantum-algebra)
based structure applied \underline{in the fermion sector}.

It remains to recall that all our treatment in secs.4-7
using the $q$-algebras $U_q(su_n)$ concerned just the fermion sector
although at the level of baryons as 3-quark bound states of fundamental
fermions. Hence, it is natural to assert that there exists direct
connection of the $q$-parameter involved in (13), (14) with fermion
mixing angle. Setting
$\theta_{\bf 10}=g(\theta_{\rm\scriptscriptstyle{C}})$ and
$\theta_{\bf 8}=h(\theta_{\rm\scriptscriptstyle{C}})$ we find for
the functions $g(\theta_{\rm\scriptscriptstyle{C}})$ and
$h(\theta_{\rm\scriptscriptstyle{C}})$ remarkably
simple explicit form:
\begin{equation}
\theta_{\bf 10}=\theta_{\rm\scriptscriptstyle{C}},
\hspace{26mm}  \theta_{\bf 8}=2~\theta_{\rm\scriptscriptstyle{C}}.              \label{(31)}
\end{equation}
With $\theta_{\bf 8}=\frac{\pi}{7}$ (see (11))
this suggests for Cabibbo angle the exact value
$\frac{\pi}{14}$.

\section{Discussion}
Quantum groups and their Hopf dual counterpart - quantum
universal enveloping algebras (QUEA) incorporate
transformation/covariance properties of related quantum vector
spaces \cite{gavr25}.  In the context of quantum homogeneous spaces
(see e.g., \cite{gavr26}) the corresponding quantum qroups act (say,
on their noncommuting 'coordinates') in a nonlinear way, as it was
exemplified  \cite{gavr27} with quantum ${CP_q}^n$.
Both quantum groups and their dual QUEA provide necessary tools in
constructing \cite{gavr28,gavr17} covariant differential calculi
and particular noncommutative geometry on quantum spaces.

In the case at hand the {\em internal} symmetries, underlying our
treatment of baryon mass sum rules in secs. 4-7 and based on the
broken $U_q(su_n)$ $(n\ge 3)$ as well as unbroken isospin $U_q(su_2)$
$q$-algebras, are closely related to certain internal or extra
(as regards the Minkowski space $M^{3,1}$) spacetime dimensions.
From this we infer the following. The above justified direct link
(31) between the Cabibbo angle
$\theta_{\rm\scriptscriptstyle{C}}=\frac{\pi}{14}$ and
the $q$-parameter, which measures strength of $q$-deformation for
the $q$-algebras $U_q(su_n)$ of flavor symmetry, can be
viewed as an indication of noncommutative-geometric origin of
fermion mixing.  In this context, the value
$\theta_{\rm\scriptscriptstyle{C}}=\frac{\pi}{14}$ of the Cabibbo angle
would serve as the noncommutativity measure of relevant quantum space
(responsible for the mixing and explicitly as yet unknown)
in extra dimensions. Concerning the latter, one can assert
that their number is not less than 2.

\vspace{2mm}
\noindent
{\bf Acknowledgements.}
I would like to thank the organizers for creating stimulating and
warm atmosphere at this NATO workshop. The research contained in
this paper was supported in part by Award No. UP1-2115 of the U.S.
Civilian Research and Development Foundation for the Independent
States of the Former Soviet Union (CRDF).


\end{document}